\documentclass[12pt]{article}
\usepackage{epsfig}
\begin{document}

\begin{center}{\bf CALCULATION OF THE CASIMIR FORCE BETWEEN
SIMILAR AND DISSIMILAR METAL PLATES AT FINITE TEMPERATURE}

\vspace{1cm}
 V. S. Bentsen, R. Herikstad, S. Skriudalen, I.
Brevik\footnote{E-mail: iver.h.brevik@ntnu.no; corresponding
author.}

\bigskip

Department of Energy and Process Engineering, Norwegian University
of Science and Technology, N-7491 Trondheim, Norway

\bigskip

and

\bigskip

J. S. H{\o}ye\footnote{E-mail: johan.hoye@phys.ntnu.no}

\bigskip

Department of Physics, Norwegian University of Science and
Technology, N-7491 Trondheim, Norway

\begin{abstract}
The Casimir pressure is calculated between parallel metal plates,
containing  the materials Au, Cu, or Al. Our motivation for making
this calculation is the need of comparing theoretical predictions,
based on the Lifshitz formula, with experiments that are becoming
gradually more accurate. In particular, the finite temperature
correction is considered, in view of the recent discussion in the
literature on this point. A special attention is given to the case
where the difference between the Casimir pressures at two
different temperatures, $T=300$ K and $T=350$ K, is involved. This
seems to be a case that will be experimentally attainable in the
near future, and it will be a critical test of the temperature
correction.
\end{abstract}
Revised version, September 2005

\end{center}

PACS numbers: 03.70.+k, 12.20.-m, 42.50.Pq

\section{Introduction}

The Casimir effect \cite{casimir48} has in recent years attracted
a great deal of interest (for recent reviews, see
\cite{milton04,milton01,bordag01,lamoreaux05}). The advent of
accurate experiments has accentuated the need of performing
detailed calculations of the Casimir forces, based upon realistic
input values for the permittivities in the (assumed homogeneous)
materials. In the case of two semi-infinite media separated by a
gap $a$ - the standard set-up in the Casimir context - the formula
in question is that due to  Lifshitz \cite{lifshitz56}. In the
case of a micrometer-sized sphere above a plane substrate - a case
that is tractable via use of the proximity force approximation
when the spherical surface is weakly curved \cite{blocki77} - the
 experimental accuracy is claimed in the literature to be
on the 1 \% level. We shall not here  give an overview of recent
experiments; we will return to some examples below. The reader may
instead consult recent reviews: a
   detailed exposition on the experiments up to 2001 is given
 in   Bordag {\it et al.} \cite{bordag01},  a  survey of
 the developments in the last four years is given by Milton \cite{milton04}, section
 3.6, and the works of the Purdue group is presented by Decca {\it
 et al.} \cite{decca05}. (A brief survey of the experiments is
 given also in the note \cite{brevik05}.)

\begin{figure}[t]
\begin{center}
\epsfig{figure=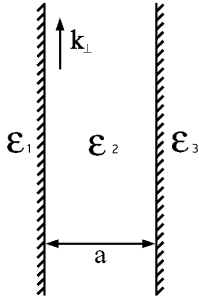, width=0.2\textwidth}
 \caption{Parallel plate-geometry.
The two surfaces are located at $z=0$ and $z=a$.} \label{geometry}
\end{center}
\end{figure}

 We shall consider the simple set-up shown in figure \ref{geometry}. There are
 two metallic semi-infinite media of permittivities
 $\varepsilon_1$ and $\varepsilon_3$, with a dielectric medium of
 permittivity $\varepsilon_2$ in between. For simplicity we assume
 that region 2 is vacuum (air), so that $\varepsilon_2=1$. The
 surfaces are assumed to be perfectly flat,  of infinite
 extension, and  the media are assumed nonmagnetic. Our intention is to
 work out  values for the attractive Casimir surface
 pressure ${\cal F}(a)$ versus gap width $a$ for similar and dissimilar metals,
   when the temperature
 $T$ is finite. Of main interest will be the temperature
 correction, in view of the conflicting opinions in the literature
 on this point. We will follow the same calculational strategy as
 in our earlier recent papers on these issues
 \cite{hoye03,brevik05a,brevik04a}.

 We shall consider three different metals; gold, copper, and
 aluminium. For these metals we have access to excellent numerical
 data for the permittivities (courtesy of Astrid Lambrecht and
 Serge Reynaud). We know how $\varepsilon(i\zeta)$ varies with
 imaginary frequency $\zeta$ over seven decades, $\zeta \in
 [10^{11}, 10^{18}]$ rad/s, at room temperature. For frequencies
 up to about $1.5\times 10^{15}$ rad/s the data are nicely
 reproduced by the Drude dispersion relation
 \begin{equation}
 \varepsilon(i\zeta)=1+\frac{\omega_p^2}{\zeta(\zeta+\nu)},
 \label{1}
 \end{equation}
 where $\omega_p$ is the plasma frequency and $\nu$  the relaxation frequency.
 For the three metals mentioned we have \cite{lambrecht00,decca03}
 \[ \omega_p= 9.0\,{\rm eV}, \quad \nu=35\,{\rm meV}\quad {\rm Au} \]
 \[ \omega_p=9.05\, {\rm eV}, \quad \nu=30\, {\rm meV} \quad {\rm Cu}\]
 \begin{equation}
 \omega_p=11.5\,{\rm eV}, \quad \nu=50\, {\rm meV} \quad {\rm
 Al}\label{2}
 \end{equation}
(note that 1 eV=$1.519\times 10^{15}$ rad/s). Using these data we
can calculate the Casimir pressures to an accuracy better than
1\%.

We shall consider three different temperatures. First, it is of
interest to work out explicitly the zero-temperature Casimir
pressure. When discussing finite temperature corrections one
should first of all know what is meant by the $T=0$ reference
level. This issue is not trivial, since most of the $T=0$
theoretical predictions have been referring to the idealized case where
 $\varepsilon=\infty$ from the outset. As discussed extensively in earlier works
\cite{milton04,hoye03,brevik04a,brevik05a}, the correct model in
an idealized setting is the modified ideal metal (MIM) model,
which assumes unit reflection coefficients for all but the
transverse electric (TE) zero frequency mode. Our argument rests
upon the condition that the relaxation frequency
 $\nu(0)$ at zero frequency remains different from zero. Here,  as in Ref.~\cite{brevik05a},
  we will calculate the $T \approx 0$ pressure numerically, inserting real data for $\varepsilon(i\zeta)$.
   We shall choose $T=1$ K as the lower temperature limit. It turns out that this limit is stable numerically,
   and numerical trials around this limit indicate that it describes the zero temperature case with good accuracy.
 This method, although numerically
demanding,  is physically better than adopting the simple
idealized metal model.

The second temperature of interest is room temperature, $T=300$ K.
Recognizing that the difference between Casimir pressures at $T=0$
and $T=300$ K will hardly become a measurable quantity we shall
instead consider, as our third chosen temperature,  $T=350$ K. The
difference between the Casimir pressures at the two last-mentioned
temperatures will perhaps soon become accessible in experiment. We
shall therefore focus upon calculating how this difference varies
with $a$.

The present calculations, involving three different metals and
three different temperatures, are of interest for comparison of
future experimental work on the finite temperature Casimir force.
To our knowledge, these are the first calculated results for
unequal metal surfaces at finite temperature. We ought here to
mention that the first correct calculations of finite temperature
Casimir forces between two {\it equal} gold, copper, or aluminium
metal half planes were made by Bostr{\"o}m and Sernelius
\cite{bostrom00,sernelius01}; these authors considered also two
thin  equal metal films of gold, silver, copper, beryllium, or
tungsten \cite{bostrom00a,bostrom00b}. The calculations by
Bostr{\"o}m and Sernelius showed that the retarded van der Waals
or Casimir interaction in more or less the same separation range
(around 1 $\mu$m) as considered in the present paper depends on
the choice of material for two equal metal surfaces. The present
work is thus a step beyond this in that it considers unequal metal
surfaces.

It has repeatedly been pointed out by some authors   that our use
of the Drude dispersion relation runs into conflict with the
Nernst theorem in thermodynamics; cf., for instance,
\cite{bezerra05,bezerra04}. We have shown earlier, however, that
this is not the case \cite{brevik05a,hoye03}. Thus, these
thermodynamic issues will not be given further attention here.

In the next section we present the general formalism, for similar as well as for dissimilar media, and give then in section 3 the results of our calculations in several diagrams. With respect to the temperature correction, we restrict ourselves in this section to the difference between $T=0$ and $T=300$ K predictions. In section 4 we focus our attention on the difference between the mentioned 300 K and 350 K cases.

In the main text, we put $\hbar =c=k_B=1$.

\section{Basic formalism}

We consider first the case of two identical media,
$\varepsilon_1=\varepsilon_3 \equiv \varepsilon$. With the same
notation as in \cite{hoye03,brevik05a} we can write the Casimir
pressure as
\begin{equation}
{\cal F}=-\frac{1}{\pi \beta
a^3}{\sum_{m=0}^\infty}'\int_{m\gamma}^\infty y^2dy
\left[\frac{A_me^{-2y}}{1-A_me^{-2y}}+\frac{B_me^{-2y}}{1-B_me^{-2y}}\right],
\label{3}
\end{equation}
where
\[ A_m=\left(\frac{\varepsilon p-s}{\varepsilon p+s}\right)^2,
\quad B_m=\left(\frac{s-p}{s+p}\right)^2,\quad s=\sqrt{\varepsilon
-1+p^2},\]
\[ p=\frac{q}{\zeta_m}, \quad y=qa,\quad q=\sqrt{k_\perp^2+\zeta_m^2}, \]
\begin{equation}
\zeta_m=\frac{2\pi m}{\beta},\quad \gamma=\frac{2\pi a}{\beta}.
\label{4}
\end{equation}
The prime on the summation sign means that the $m=0$ term is
counted with half weight; $\beta=1/T$ is the inverse temperature.
The minus sign in Eq. ({\ref{3}) means that the force is
attractive.

[The following point should be noted. Assume that a plane wave is
incident from the left medium ($z<0$) towards the boundary at
$z=0$. For the TM mode, the ratio between the reflected wave
amplitude $R^{TM}$ and the incident wave amplitude $A^{TM}$ is
equal to the square root of the coefficient $A_m$, after the real
frequency $\omega$ has been replaced with the imaginary frequency
$\zeta$ ($\omega=i\zeta$), corresponding to the surface mode:
\begin{equation}
\frac{R^{TM}}{A^{TM}}=\sqrt{A_m} \,; \label{5}
\end{equation}
cf. Appendix B in \cite{hoye03}. Analogously, for the TE mode:
\begin{equation}
\frac{R^{TE}}{A^{TE}}=\sqrt{B_m}\,.] \label{6}
\end{equation}
Generalization to the case of dissimilar media leads to the
following expression:
\begin{equation}
{\cal F}=-\frac{1}{\pi \beta
a^3}{\sum_{m=0}^\infty}'\int_{m\gamma}^\infty y^2dy \left[
\frac{\Delta_1^{TM}\Delta_2^{TM}e^{-2y}}
{1-\Delta_1^{TM}\Delta_2^{TM}e^{-2y}}+\frac{\Delta_1^{TE}\Delta_2^{TE}e^{-2y}}
{1-\Delta_1^{TE}\Delta_2^{TE}e^{-2y}}\right],\label{7}
\end{equation}
where
\[ \Delta_1^{TE}=\frac{s_1-p}{s_1+p},\quad
\Delta_2^{TE}=\frac{s_3-p}{s_3+p},\]
\begin{equation}
\Delta_1^{TM}=\frac{\varepsilon_1p-s_1}{\varepsilon_1p+s_1},\quad
\Delta_2^{TM}=\frac{\varepsilon_3p-s_3}{\varepsilon_3p+s_3}.
\label{8}
\end{equation}
Again, if the two media are equal we have $\Delta_1=\Delta_2$ for
each of the modes, so that
\begin{equation}
\Delta_1^{TM}\Delta_2^{TM} \rightarrow A_m,\quad
\Delta_1^{TE}\Delta_2^{TE} \rightarrow B_m, \label{9}
\end{equation}
and the formula (\ref{3}) is recovered.

We  calculate the expression (\ref{7}) by means of MATLAB. The
zero frequency case $m \rightarrow 0$ may be treated separately by
analytical methods, at least if we are considering an idealized
model for the metal, because in this limit there is  an interplay
with the other limit $\varepsilon \rightarrow \infty$ in the
expressions for the coefficients. We have in this case
$p\rightarrow \infty,\, s_i\rightarrow p$ for $i=1,3$, implying
that
\begin{equation}
\Delta_i^{TE}\rightarrow 0, \quad \Delta_i^{TM}\rightarrow
\frac{\varepsilon_i-1}{\varepsilon_i+1} \equiv \Delta_i.
\label{10}
\end{equation}
Then, the $m=0$ contribution can be written as
\begin{equation}
{\cal F}_0=\frac{1}{\pi \beta a^3}I_0, \label{11}
\end{equation}
where
\begin{equation}
I_0=-\frac{1}{2}\int_0^\infty y^2dy \frac{\Delta e^{-2y}}{1-\Delta
e^{-2y}}= -\frac{1}{8}{\rm polylog}(3,\Delta), \label{12}
\end{equation}
the polylog function being defined as ${\rm polylog}
(x,z)=\sum_{n=1}^\infty z^n n^{-x}$. We here assume that $\Delta
\equiv \Delta_1\Delta_2 \leq 1$. (For $\Delta>1$ the integral is
undefined.) Since $\varepsilon_i \gg 1$ for a metal near $\zeta=0$
we have that $\Delta \approx 1$, but still less than unity, so we
can let $I_0\rightarrow -\frac{1}{8}{\rm
polylog}(3,1)=-\frac{1}{8}\zeta(3)$, where $\zeta(x)$ is the
Riemann zeta function. The numerical value of $I_0$ used in our
calculations was $I_0=-0.1502571129.$

The first equation in (\ref{10}) means that there is no
contribution to the Casimir force from the $m=0$ TE mode, at
finite temperatures. (At $T=0$ the $m=0$ effect vanishes, as the
discrete  Matsubara  sum is replaced by an integral over
frequencies.) This behaviour is a consequence of the Drude
relation at low frequencies. The same behaviour can also be seen
by use of quantum statistical methods, as was
demonstrated for the case of spherical geometry in
\cite{hoye01,brevik02}.

\section{Numerical calculations. Results}

Since expression (\ref{7}) is complicated, with an upper limit
$y=qa =\infty$ for the integral, it is useful first to get
information about how the integrand varies with respect to $y$  in
typical cases. The expression is most demanding numerically for
low temperatures and small gap widths.


\begin{figure}[t]
  \begin{center}
    \epsfig{figure=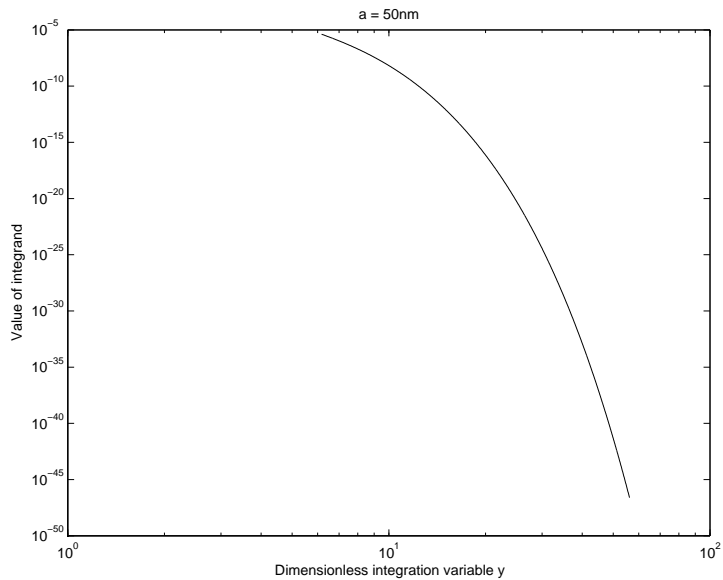, width=0.7\textwidth}
    \caption{Integrand of equation~(\ref{7}) versus $y=qa$ for $a=50$ nm,
         $m=45000$, $T=1$ K. The materials are Al and Cu. Left end point of the
          curve corresponds to the lower limit $y=m\gamma$ of the integral in (\ref{3}).}
    \label{highMval}
  \end{center}
\end{figure}

\begin{figure}[h]
  \begin{center}
    \epsfig{figure=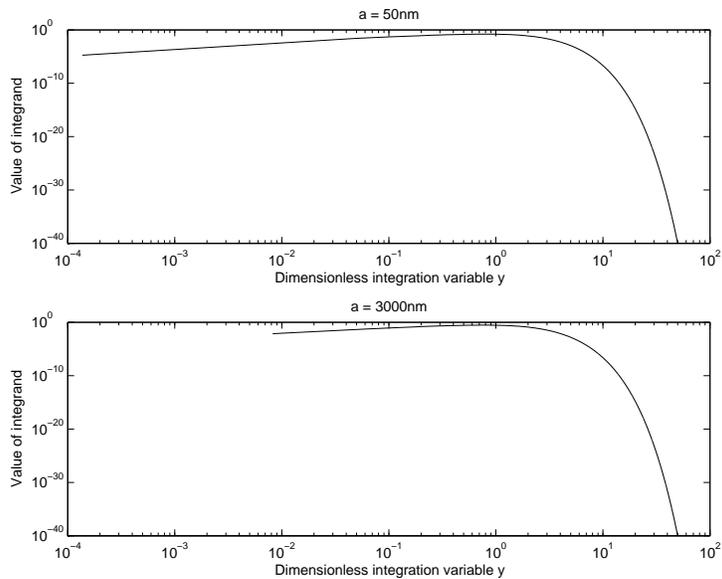,width=0.7\textwidth}
    \caption{Same as  figure \ref{highMval}, but with $a=\{50,3000\}$ nm, and $m=1$.}
    \label{lowMval}
  \end{center}
\end{figure}


Figure \ref{highMval} shows how the
integrand varies with respect to $y$ for a very high Matsubara
number, $m=45000$, when $a=50$ nm and $T=1$ K. The configuration is
one aluminium and one copper plate. It is seen that when $y$
becomes larger than about 10, the contribution to the integral
decreases rapidly. At $y=50$ the value is only $10^{-45}$. Figure
\ref{lowMval} shows for comparison how the integrand varies with $y$ at the
same temperature when the frequency is at the lowest non-vanishing
value, $m=1$, for $a=50$ nm and $a=3000$ nm. The behaviour is seen
to be quite similar to that above; for instance, when $y=50$ the
integrand becomes approximatively $10^{-40}$. For different
plate combinations, we get approximately the
same behaviour.   For high temperatures, the same
conclusion can be drawn. In all, we found it sufficient in our
computations to adopt the value
\begin{equation}
y_{max}=50 +m\gamma \label{13}
\end{equation}
as general cutoff. [It is useful to note that $\gamma =2744
\,(aT)$, when $a$ is given in meters and $T$ in degrees kelvin.]

Numerically, we used a method of higher order recursive adaptive
quadrature. This method approximates the value of the integral
with a chosen tolerance of $10^{-10}$.

\begin{figure}[h]
  \begin{center}
    \epsfig{figure=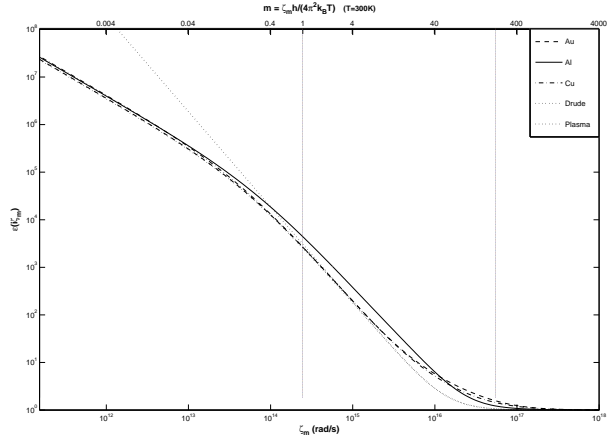,width=0.7\textwidth}
    \caption{Numerical permittivity data for
Al, Au and Cu (courtesy of Astrid Lambrecht and Serge Reynaud).
Vertical lines show the frequency region that we used in the $m$
summation at $T=300$ K. Predictions from the Drude and the plasma
dispersion relations are also shown in  the case of Au. Top axis
gives the values of the Matsubara number $m$.}
    \label{disprel1}
  \end{center}
\end{figure}
\begin{figure}[h]
  \begin{center}
    \epsfig{figure=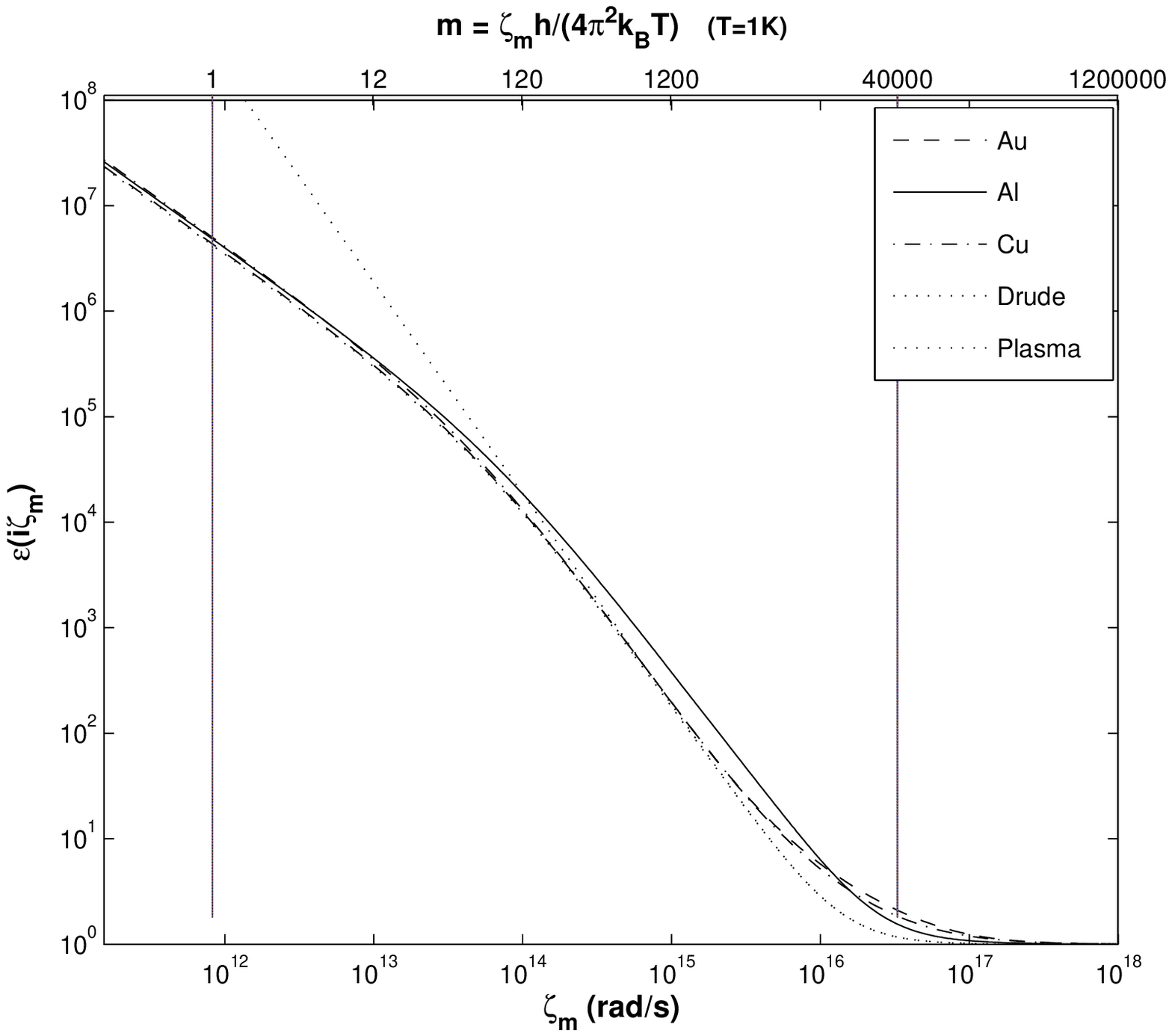,width=0.7\textwidth}
    \caption{Same room temperature data as
in figure \ref{disprel1}, but the vertical lines show the
frequency region used in the calculations  at $T=1$ K. Drude and
plasma dispersion relations shown for Au as before.}
    \label{disprel2}
  \end{center}
\end{figure}

Next, it is useful to show the variations of $\varepsilon(i
\zeta)$ graphically for the three metals mentioned, together with
information about the frequency region actually used in the
calculations. Figure \ref{disprel1} shows this for the case of
$T=300$ K. The data extend over 7 decades. The vertical lines show
that the important frequency region in this case lies between
$2\times 10^{14}$ rad/s and $5.5\times 10^{16}$ rad/s. We also
show the corresponding values of $m=\zeta_m \beta/2\pi$ (in
nondimensional units). The range of $m$ is [1,220]. For
comparison, the theoretical predictions are  shown for the case of
gold, both when using the Drude relation (\ref{1}) and when using
the plasma dispersion relation
\begin{equation}
\varepsilon(i \zeta)=1+\frac{\omega_p^2}{\zeta^2}. \label{14}
\end{equation}
We see that for $\zeta < 1.5\times 10^{15}$ rad/s the Drude curve
fits the data nicely, but for $\zeta > 2\times 10^{15}$ rad/s it
gives too low values for $\varepsilon$. The used frequency region
corresponds to the area where the Drude prediction and the plasma
prediction are approximately equal, and also to the area where the
data for aluminium differ the most from the data for gold and
copper.

Figure \ref{disprel2} shows the analogous situation when $T=1$ K. It becomes
now necessary to use a much larger frequency region, from $7\times
10^{11}$ rad/s to $3.3\times 10^{16}$ rad/s, corresponding to
about 80\% of the entire data set. The Matsubara number region is
 $m\in [1,40000]$. The physical reason for this behaviour is, as always,
 that the case of low temperatures implies that the frequencies
 are very closely spaced.

 \subsection{Room-temperature Casimir force}

 We shall in this subsection assume that $T=300$ K. The force,
 expression (\ref{7}), is always negative, but it is convenient to
 represent it graphically in terms of the modulus $|{\cal F}(a)|$.
 Since in the parallel plate-experiment of Bressi {\it et al.}
 \cite{bressi02} one was able to control the gap width down to
 50 nm, we choose $a=50$ nm as our lower limit. As upper limit we
 choose $a=3\,\mu$m. To better visualize the results graphically,
 we divide all data sets into two groups, $a\in [50,200]$ nm and
 $a\in [200,3000]$ nm. As mentioned earlier, the plates are assumed
  infinite, and all roughness corrections
 are ignored.

\clearpage
\begin{figure}[!h]
  \begin{center}
    \includegraphics[angle=90, width=\textwidth]{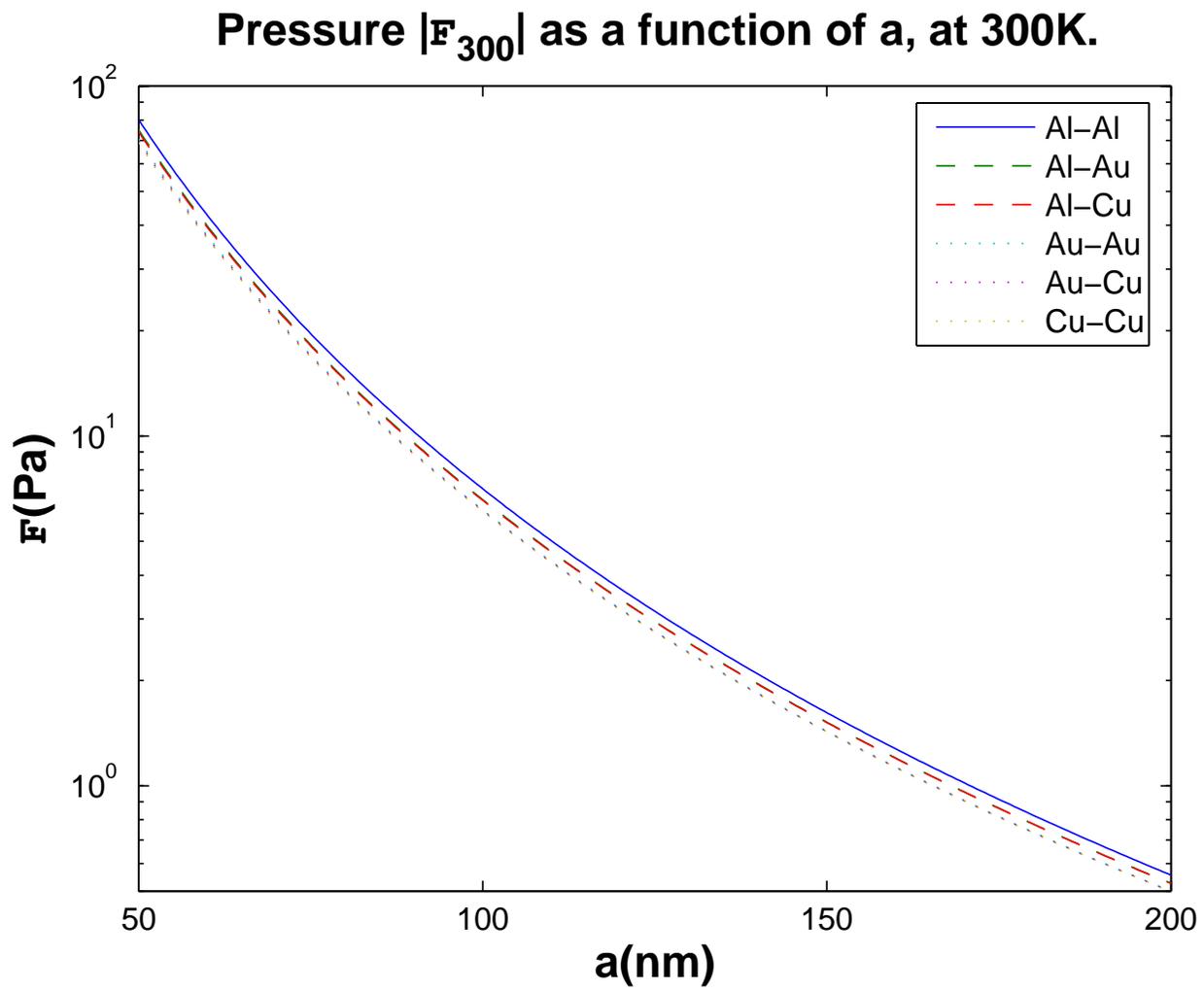}
    \caption{Casimir pressure for various
combinations of metal plates, versus gap width $a\in [50,200]$ nm
when $T=300$ K.}
    \label{fa300short}
  \end{center}
\end{figure}

\begin{figure}[!h]
  \begin{center}
    \includegraphics[angle=90, width=\textwidth]{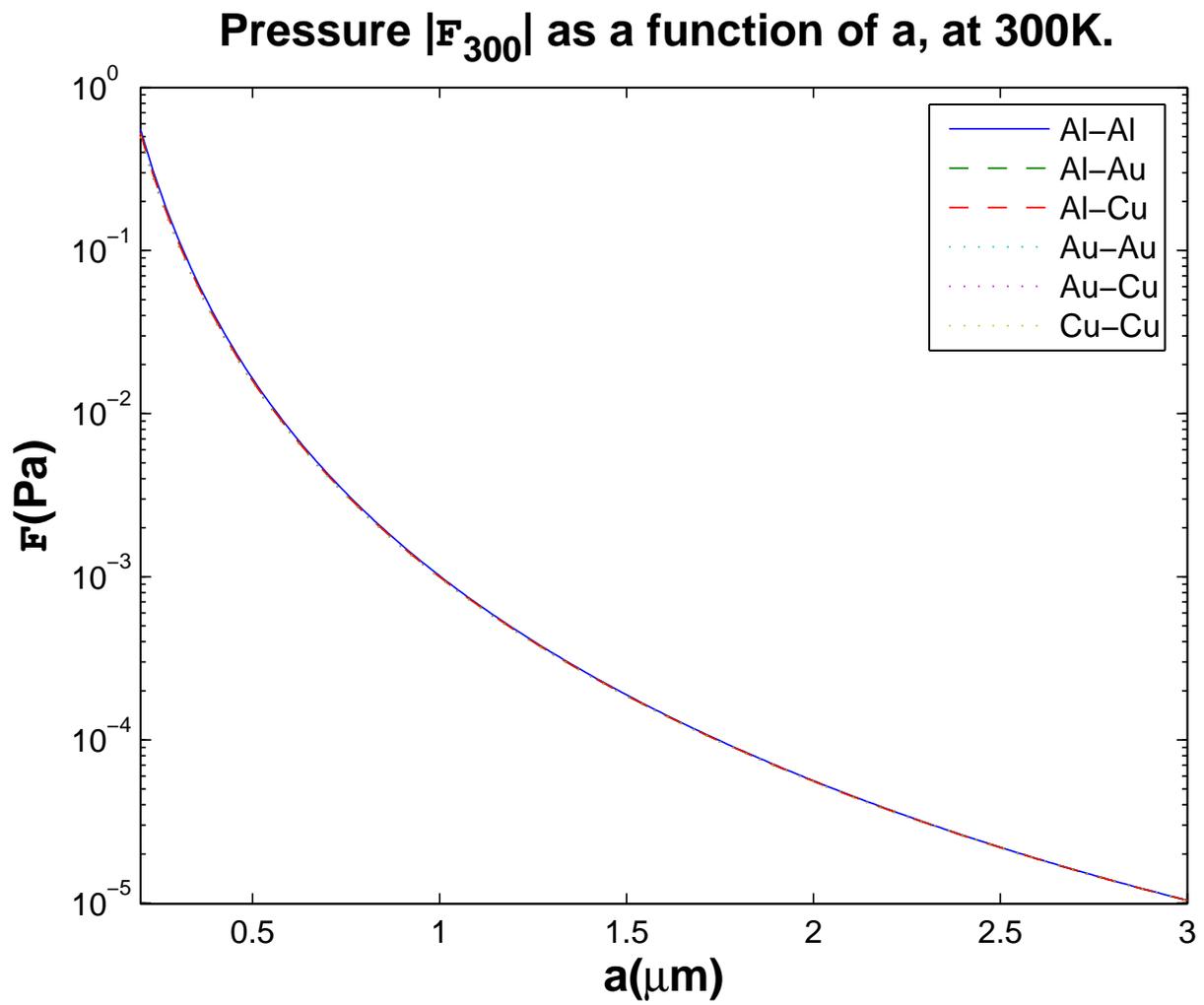}
    \caption{Same as figure \ref{fa300short}, but with
$a\in [200,3000]$ nm when $T=300$ K.}
    \label{fa300long}
  \end{center}
\end{figure}
\clearpage

 Figures \ref{fa300short} and \ref{fa300long} show how the Casimir pressure varies with $a$,
 for various combinations of metal plates. The force between two
 gold plates was computed earlier in \cite{brevik04a}, but is
 included here for comparison. The differences between the various
 combinations of the materials are seen to be small, and they
 diminish with increasing gap widths. The largest force always occurs for
 two aluminium plates. This may be called group I. The combinations aluminium-gold and
 aluminium-copper yield a somewhat smaller force (group II), and the last
 combinations Au-Au, Au-Cu and Cu-Cu result in the weakest set (group III).
 When $a$ increases from small to large values, the internal order
 in strength between the materials in groups II and III  is interchanged.

 \subsection{Room-temperature correction, compared to $T=0$}

 We now turn to the finite temperature correction in the Casimir
 force. As mentioned above, it is then important  to make clear what we mean
 with the zero-temperature force. Numerically, we have seen that
   it becomes satisfactory to represent the  latter case by the choice
  $T=1$ K, the difference between $T=0$ and $T=1$ K
 being negligible.

 Since the data for $\varepsilon(i\zeta)$ are measured at room
 temperature, the natural question becomes: can we use these data
 also at very low temperatures? This issue has been discussed
 earlier, in \cite{brevik05a,brevik04a,bostrom04}, with the
 conclusion that the temperature dependence appears not to
 influence the dispersion relation in a way that changes the
 Casimir force significantly. This implies that we can insert
 the same Lambrecht-Reynaud data as before, and also assume
 the same values for $\omega_p$ and $\nu$  (cf. equation
 (\ref{2})).

The calculated results from the $T=1$ K case are hardly
distinguishable from the $T=300$ K case when plotted in the same
figure. Therefore, it is better to show the calculated
 finite-temperature {\it differences}. In figures \ref{tempDif1} and \ref{tempDif2} we
 show the difference between the Casimir pressures at $T=1$ K and
 $T=300$ K,
 \begin{equation}
 \Delta {\cal F}=|{\cal F}_1| - |{\cal F}_{300}|, \label{15}
 \end{equation}
 versus $a$ in the range from 50 nm to 1700 nm. As the graphs for
 the various combinations of metals become indistinguishable for
 $a>0.4\,\mu$m and approach zero when $a$ becomes larger than
 $1.7\,\mu$m, we have omitted the region from $1.7\,\mu$m to
 $3\,\mu$m in figure \ref{tempDif2}.

\clearpage
\begin{figure}[!h]
  \begin{center}
    \includegraphics[angle=90, width=\textwidth, height=\textheight]{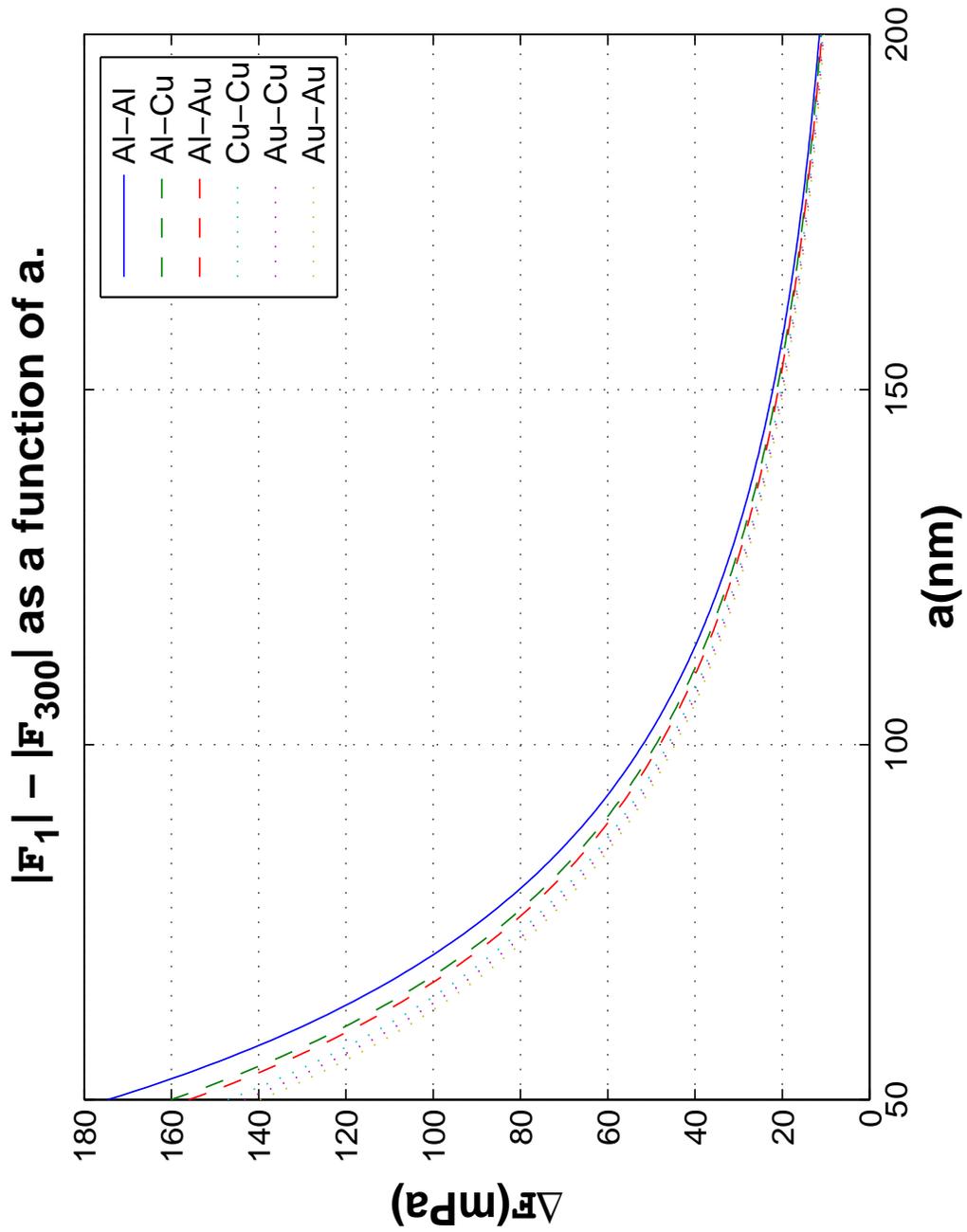}
    \caption{Difference between the Casimir
pressures at $T=1$ K and $T=300$ K, equation (\ref{15}), versus
gap width $a$ for  $a\in [50,200]$ nm.}
    \label{tempDif1}
  \end{center}
\end{figure}

\begin{figure}[!h]
  \begin{center}
    \includegraphics[angle=90, width=\textwidth, height=\textheight]{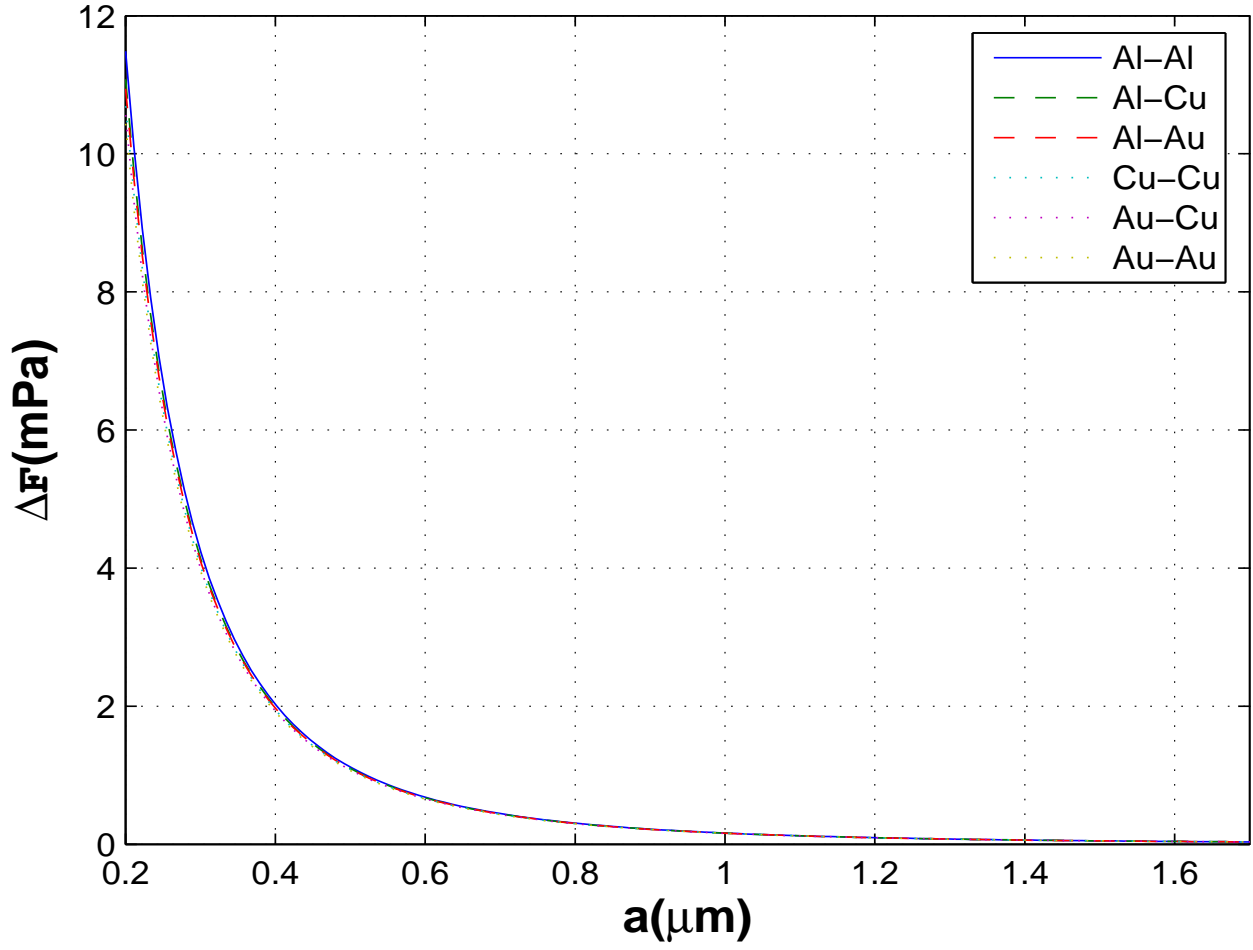}
    \caption{Same as figure \ref{tempDif1}, for the
interval $a\in [200,1700]$ nm.}
    \label{tempDif2}
  \end{center}
\end{figure}
\clearpage

 An important property seen from the curves is that $\Delta \cal F$ is
 positive. That means, the force is weaker at room temperature
 than at $T=0$. This is the same effect as  we have pointed out
 earlier, in connection with identical materials in the plates
 \cite{brevik04a,hoye03,brevik05a}; see also \cite{milton04,sernelius04}.
  The behaviour is a direct
 consequence of the Lifshitz formula in combination with realistic
 permittivity data for the materials, the latter being, as we have seen,  in
 agreement with the Drude relation at frequencies $\zeta < 1.5\times 10^{15}$ rad/s. Our
 results for the temperature dependence are in contrast to those
 obtained by use of the plasma dispersion relation; in that case,
 the deviation of the force is positive instead of negative, and
 is moreover very small \cite{decca03,chen03,chen04}.

 To get an overview of the magnitudes of the temperature
 correction, let us give some examples:

 1.  For small gap widths, the correction is relatively small.
 Thus when $a=100$ nm, the Casimir pressure is 6.105 Pa at $T=1$ K
 and 6.061 Pa at $T=300$ K, thus giving a room temperature
 reduction of 0.72\%.

 2.  When $a=200$ nm, the respective pressures are 510 mPa at 1 K
 and 500 mPa at 300 K, giving a 2\% reduction.

 3.  When $a=500$ nm, the pressures are 16.3 mPa and 15.2 mPa,
 giving a 6.7\% reduction.

 4.  For large gap widths the percentage corrections become much
 higher, though the pressures themselves are much weaker  thus
  making experimental work more difficult.
 When $a=1\,\mu$m the respective pressures are 1.12 mPa and 0.96
 mPa, giving a 13.9\% reduction.

 It is of interest to compare the above findings with figure \ref{disprel2} in
 \cite{hoye03}. That figure shows, in the case of Au-Au, how the
 Casimir pressure varies with $aT$ when the media are assumed {\it
 nondispersive}.   The width is assumed to be $a=1\,\mu$m. The case temperature $T=300$ K corresponds to
 $aT=0.131.$ One sees that in this case the agreement with our
 result above, ${\cal F}=0.96$ mPa, is reasonably good, if we put $\varepsilon =3000$.  The broken
 line in figure \ref{disprel2} in \cite{hoye03} gives the result when the modified ideal metal
 model  is used in the calculation. As already mentioned, this is an idealized
 model, which assumes unit reflection coefficients for all but the
 TE zero mode:
 \begin{equation}
 A_0=1, \quad B_0=0, \quad A_m=B_m=1\,\, {\rm for}\,\, m \geq 1.
 \label{16}
 \end{equation}
 The MIM model corresponds to $\varepsilon =\infty$. Putting
 $aT=0.131$ in the mentioned figure \ref{disprel2} we see that ${\cal F}
 \approx 1.1$ mPa. There is thus more than a 10\% overprediction
 of the Casimir pressure following from the MIM model, at $a=1\mu$m and  $T=300$ K,
 in comparison with our result 0.96 mPa above.

 We recall again  that at $T=0$ there is no distinction
 between a MIM model and an "ideal metal" model (IM), for which
 $A_m=B_m=1$ for all $m\ge 0$. In the mentioned figure \ref{disprel2}, setting
 $aT=0$, we obtain for an ideal metal ${\cal F}\approx 1.3$ mPa at $a= 1\,\mu$m,  which is
 considerably larger than the value 1.12 mPa calculated above for
 $T=0$.

 \section{A temperature correction of experimental interest}

 The calculated temperature correction at $T=300$ K as compared
 with the $T=0$ case, although of fundamental interest, will
 be difficult to measure in practice.
 For practical purposes it can thus be better to focus on temperatures
that are more realistic in the laboratory. In the following, as an
example,  we calculate the difference between the Casimir
pressures at two temperatures $T_1=300$ K and $T_2=350$ K, and let
from now on $\Delta {\cal F}$ mean the pressure difference:
\begin{equation}
\Delta {\cal F}=|{\cal F}_{300}|-|{\cal F}_{350}|. \label{17}
\end{equation}
This idea of testing the Casimir force seems to go back to Chen
{\it et al.} \cite{chen03}, and was elaborated upon also in
\cite{brevik04a}.

Figures \ref{tempDif3} and \ref{tempDif4} show how the quantity
(\ref{17}) varies with $a$ in the interval $a\in [50,1400]$ nm,
for the same combinations of materials as before. Again, the
differences between the materials are seen to be small. Taking
Au-Au as an example, we see that $\Delta {\cal F} =2.0$ mPa when
$a=200$ nm. It would perhaps be possible to measure a quantity
like this. The experimental advantage we see of this kind of
experiment is that only force differences are involved, for a
given value of the gap width. Then there will be no need of
measuring the absolute Casimir pressure itself, to an extreme
accuracy. (Thermal expansion effects, of course, will have to be
taken into account.)

\clearpage
\begin{figure}[!h]
  \begin{center}
    \includegraphics[angle=90, width=\textwidth, height=\textheight]{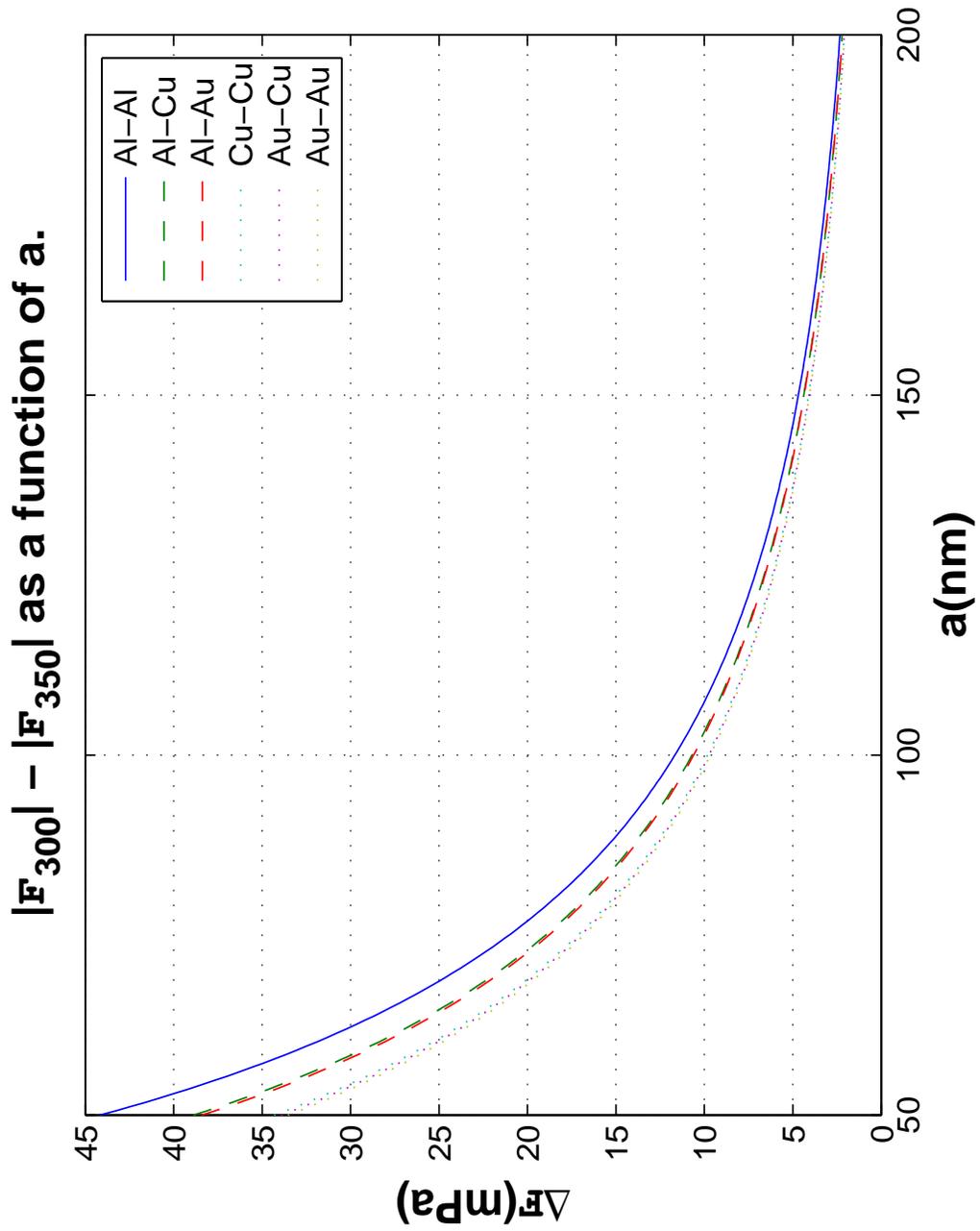}
    \caption{Difference between Casimir
pressures at $T_1=300$ K and $T_2=350$ K, equation (\ref{17}),
versus $a$ for $a\in [50,200]$ nm.}
    \label{tempDif3}
  \end{center}
\end{figure}
\begin{figure}[!h]
  \begin{center}
    \includegraphics[angle=90, width=\textwidth, height=\textheight]{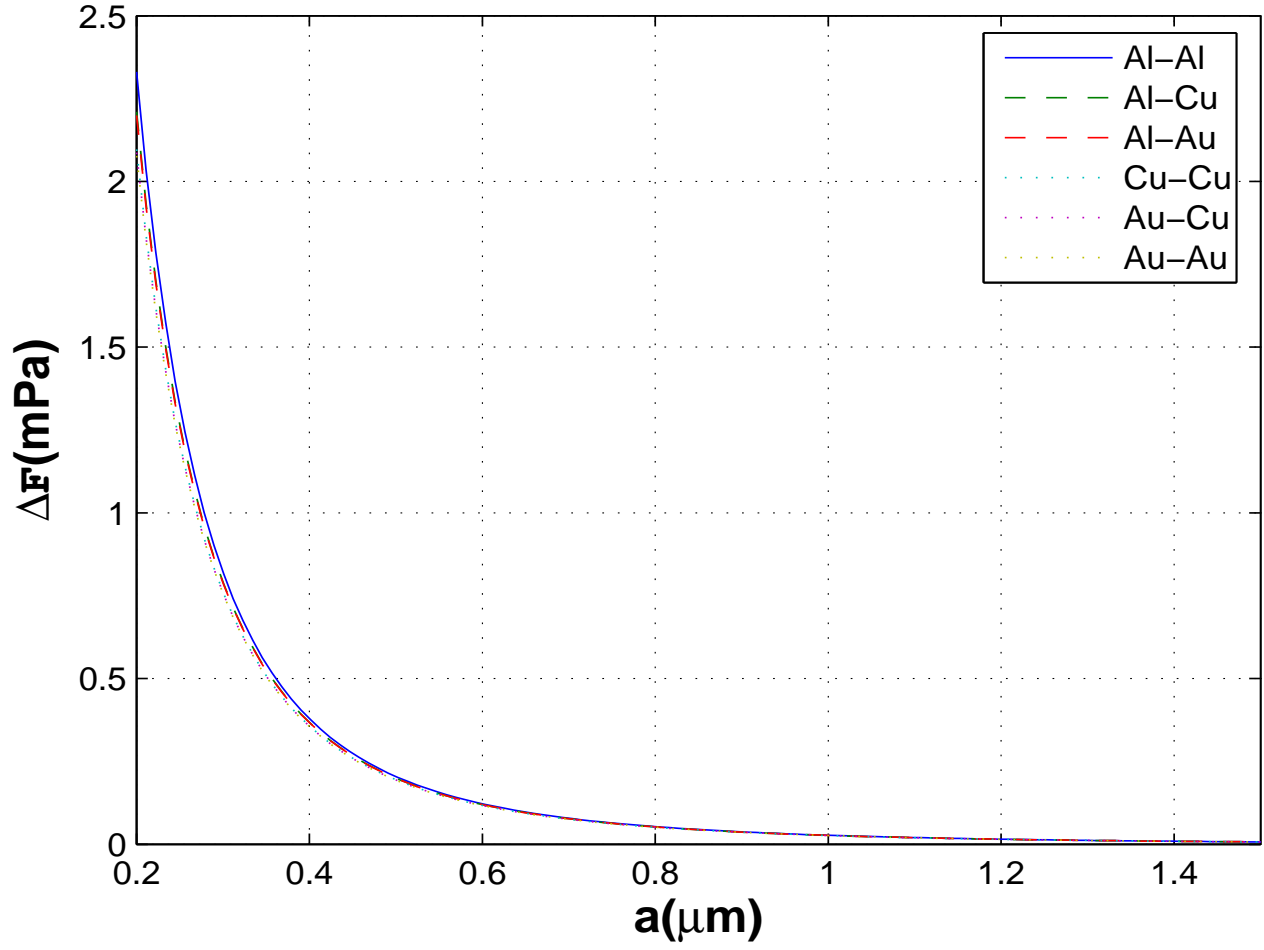}
    \caption{Same as figure \ref{tempDif3}, for the
interval $a\in [200,1400]$ nm.}
    \label{tempDif4}
  \end{center}
\end{figure}
\clearpage


Finally, figure \ref{percent} shows how the relative magnitude of the
difference Casimir pressure, $\Delta {\cal F}/{\cal F}_{300}$, for
the same two temperatures. We see, in accordance with the
behaviour above, that the relative temperature correction is
greatest when $a$ is large. When $a=1.75\,\mu$m, the correction
takes its maximum value, about 4\%. Again, the experimental
problem at large distances is that the forces themselves are so
small.


\begin{figure}[h]
  \begin{center}
    \epsfig{figure=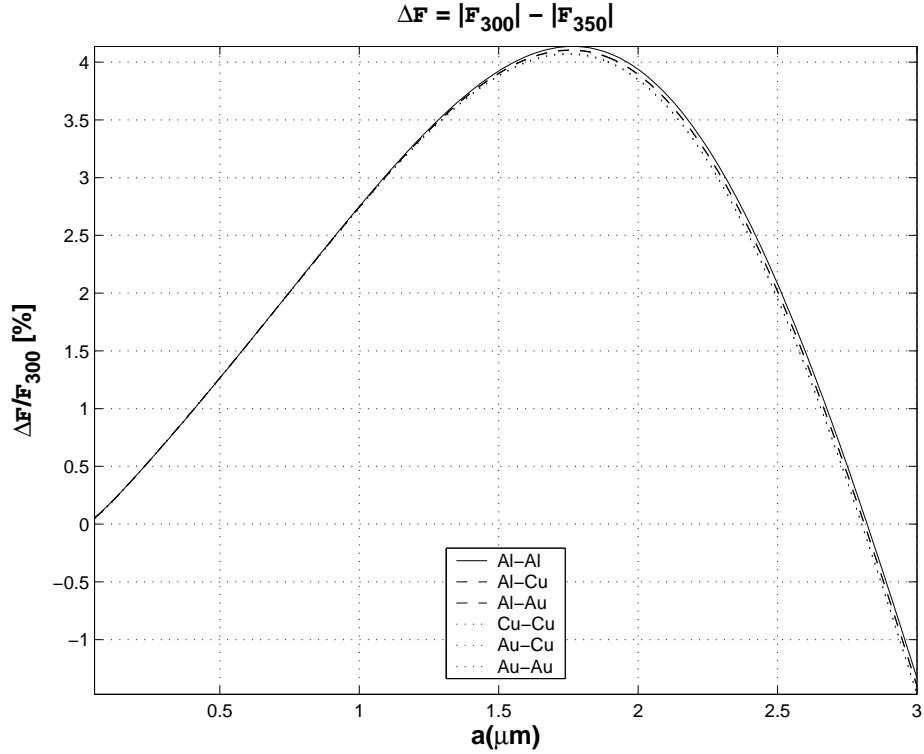, width=0.9\textwidth}
    \caption{Relative change of the
Casimir pressure, between $T_1=300$ K and $T_2=350$ K.}
    \label{percent}
  \end{center}
\end{figure}


\section{Summary}

For similar and dissimilar plates, including the metals gold, copper, and aluminium, we have made accurate calculations of the Casimir
pressure and have shown the results graphically. Basic elements in our calculations are the Lifshitz formula; cf. equations (\ref{3})
 and (\ref{7}), together with realistic room-temperature values of the permittivities $\varepsilon(i\zeta)$. For frequencies
 $\zeta <1.5 \times 10^{15}$ rad/s; cf. equations (\ref{1}) and (\ref{2}), the Drude dispersion relation is followed with great accuracy.

We show the results at three chosen temperatures: (i) at $T=1$ K, representing the $T=0$ case with good accuracy; (ii) at $T=300$ K;
 and (iii) at $T=350$ K. The low-temperature case is calculated numerically, without involving the modified ideal metal (MIM) model \cite{hoye03}.

It turns out that the differences between the Casimir pressures for the metals investigated here are small. From figures 6 and 10, for
 instance, it is seen that it is the case of Al-Al surfaces that gives the strongest forces.

The most promising option for measuring the Casimir temperature correction in practice seems to be to measure the pressure difference
between two practically accessible temperatures in the laboratory. As figure 12 shows, the relative change of the Casimir pressure between
the temperatures 300 K and 350 K are about 4\% when $a=1.75\,\mu$m. The practical problem here is that the forces themselves are so small.
For lower values of $a$ the forces increase in magnitude, but the relative temperature corrections then become smaller.

\bigskip

{\bf Acknowledgements}

\bigskip

We thank Astrid Lambrecht and Serge Reynaud for providing us with
permittivity data for Au, Cu and Al, and we thank Jan. B. Aarseth
for  help with the numerics.

\end{document}